\documentclass[useAMS,usenatbib]{mn2e}

\usepackage{graphicx}
\usepackage{times}
\usepackage{natbib}
\usepackage{color}
\newcommand{\cosmodeep}{\it{\small COSMODEEP}}

\begin{document}
 
\title{Deep Learning Based Detection of Cosmological Diffuse Radio Sources}
\author[C.Gheller, F. Vazza, A. Bonafede]{C. Gheller$^{1}$, F. Vazza$^{2,3,4}$, A. Bonafede$^{2,3,4}$\\
$^{1}$ ETHZ-CSCS, Via Trevano 131, 6900, Lugano, Switzerland \\
$^{2}$ Dipartimento di Fisica e Astronomia, Via Gobetti 93/2, 40131, Bologna, Italy\\
$^{3}$ Hamburger Sternwarte, Gojenbergsweg 112, 21029 Hamburg, Germany
$^{4}$ INAF-Istituto di Radio Astronomia, Via Gobetti 101, Bologna, Italy\\
}

\maketitle 

\begin{abstract}
In this paper we introduce a reliable, fully automated and fast algorithm to detect extended extragalactic radio sources (cluster of galaxies, filaments) in existing and forthcoming surveys  (like LOFAR and SKA).  The proposed solution is based on the adoption of a Deep Learning approach, more specifically a Convolutional Neural Network, that proved to perform outstandingly in the processing, recognition and classification of images. The challenge, in the case of radio interferometric data, is the presence of noise and the lack of a sufficiently large number of labeled images for the training. We have specifically addressed these problems and the resulting software, {\cosmodeep} proved to be an accurate, efficient and effective solution for detecting very faint sources in the simulated radio images. We present the comparison with standard source finding techniques, and discuss advantages and limitations of our new approach.
\end{abstract}

\label{firstpage} 

\section{Introduction}
\label{sec:intro}

The challenge facing  astronomers in the upcoming decade is not only scientific, but also technological. A flurry of  complex  data will be delivered by new telescopes such as SKA, LSST or CTA, and this will be difficult to manage with  traditional approaches. Data will have to be stored in dedicated facilities, providing the necessary capacity at the highest performance. Corresponding data processing will have to be performed local to the data, exploiting available high performance computing resources. Data reduction and imaging software tools will have to be adapted, if not completely re-designed, in order to efficiently run at scale. Fully automated pipelines will be a compelling requirement for effective software stacks as the  richness and complexity of incoming data  will inhibit human interaction and supervision.

In this work, we focus on radio imaging of extended and low surface brightness emission from the cosmic web \citep[e.g.][]{2011JApA...32..577B}, which may become feasible thanks to the expected ten-fold improvement in instrument sensitivity. Such large-scale diffuse and faint emission is mostly associated with the extended distribution of synchrotron emitting electrons in the largest structures of the Universe, i.e. the gas structure around galaxy clusters and filaments. This is expected to appear as  an elongated low surface brightness and flat 
spectrum radio emission ( i.e. $\alpha \sim 1$, with 
$\alpha$ being the spectral index, linked to the source flux density $S$ according to $S(\nu) \propto \nu^{- \alpha}$) tracing structure formation shocks in cluster outskirts and around cosmic filaments \citep[e.g.][]{va15radio}. Detecting this diffuse emission will be particularly important as it is expected to  carry unique information on the origin of extragalactic magnetic fields \citep[e.g.][]{va17cqg}. \\

However, identifying the faint radio signal from cosmic filaments will be particularly challenging owing to the difficulty in detecting their gas component  in any other wavelength, as well as due to the very large angular scale they typically probe (several degrees), which makes them increasingly more elusive at high radio frequencies. 
In addition, radio images obtained through interferometric observations are affected by several instrumental and environmental effects, which may increase the image noise well above the expected thermal noise threshold (e.g. radio interferometric interferences from the ground and from the sky, unstable ionospheric conditions, deconvolution artifacts). As some of these effects are direction-dependent and vary across the field of view, the final noise in the image is often non uniform and of similar level to the signal from the real sources.\\

Our goal here is to develop a source finder tool tailored to detect faint and  extended sources, with an accuracy comparable to that of the most sophisticated software available, for instance PyBDSF (see Section \ref{sec:res}), a recent Python-based tool designed for LOFAR, which is to our knowledge the most used in the field. We also require our tool to be flexible and easily extensible enough to handle different kinds of problems, for instance the analysis of multi-dimensional data, like radio data cubes.
Furthermore, it has to run efficiently on large supercomputing systems, exploiting, in particular, parallelism and accelerators, managing problems of ``any'' size at high performance. Finally, it has to be fully automated, requiring no human intervention or control, and based on portable components, in order to be usable on any computing architecture. \\

In order to develop such data processing methodology, we have explored the potential of Machine Learning, a branch of Artificial Intelligence already successfully used in astronomy (for a review see \cite{ball2009} and \cite{Kremer2017BigUB}, and for recent applications see \cite{2018MNRAS.tmp..521B}, \cite{2018arXiv180204271L}, \cite{2018arXiv180109070R}, \cite{2018MNRAS.473...38S}, \cite{2017arXiv170506818B}). Among the various Machine Learning approaches, we have focused on Deep Learning, which provides outstanding performance for tasks relating to computer vision, text analysis, fragmentation, speech recognition (\cite{726791}, \cite{NIPS2012_4824}, \cite{DBLP:journals/corr/SimonyanZ14a}, \cite{43022}, \cite{DBLP:conf/cvpr/HeZRS16}, \cite{garcia17}), among others. 
Deep Learning has become increasingly popular in the last decade thanks to two concurrent factors: the availability of enough computing power to cope with complex, multi-layered neural networks, and the availability of enough data to perform the training. Recently, it has also been adopted in applications in astronomy and cosmology (see for example \cite{2017arXiv170705167S}, \cite{2018arXiv180107615H}, \cite{2018arXiv180303084C}, \cite{2018MNRAS.476..246L}, \cite{2018MNRAS.tmp..588A}, \cite{2017arXiv170906257M}, \cite{2017arXiv170905889N}, \cite{2018MNRAS.476.1151P}, \cite{2017MNRAS.472.3101G}, \cite{2017ApJS..230...20A}, \cite{2018arXiv180106381H}).

Out of the existing Deep Learning approaches, we have focused this work on Convolutional Neural Networks (CNN), which have proved to be both efficient and accurate in classifying images. The main advantages of CNNs are their high accuracy, their high computational performance and their suitability for a broad spectrum of applications. Training, involving basic linear algebra local operations, can be performed effectively on accelerated architectures exploiting, for instance, GPUs. The network can be efficiently decomposed to run on distributed, multi-processor systems \citep{perfmod17}. Once trained, classification is a simple and fast task, and accuracy can range close to 100\% depending on the model, the task and the dataset. Furthermore, by changing a few parameters and the input data, the same model can be trained for completely different tasks. Drawbacks are represented by the lack of flexibility of a trained model, a network being designed and trained on a specific kind of data input (e.g. $2000 \times 2000$ pixel gray-scale images), and the need for {\it large, labeled} training sets. The former obviously  represents a serious concern in Astronomy due to the heterogeneity of the data products that can be delivered by different instruments, as well as due to the highly specialized format and resolution of output images from different telescopes. 

The main challenge, however, is represented by the availability of sufficiently big datasets with pre-classified (labeled) images that can be used for the training. Tens of thousands of labeled images should be accessible in order to effectively train the network. Currently, few surveys of extragalactic objects have a large enough dataset of labeled images, which makes any application of Deep Learning challenging. We have specifically addressed this problem by creating ``mock'' observations, starting from the results produced by cosmological numerical simulations (see Section \ref{sec:training}). This allows us to generate enough images to train of the network. Having the full control of the training images, we had the capability to develop a labeling algorithm able to classify and label images without human supervision.\\

The CNN based algorithm we present in this paper, called {\cosmodeep}, represents the first step towards a fully automated software pipeline able to face the challenges posed by big, complex radio data. {\cosmodeep} can not only train the CNN and classify images, but takes care of the preprocessing and labeling of the images used for the training. Hence it provides all the tools to develop an effective classification algorithm built on the top of a Deep Learning model. 

The paper is organized as follows. The details of the {\cosmodeep} CNN are presented in Section \ref{sec:cnn}. Section \ref{sec:training} focuses on the training data and how images are generated, labeled, and processed, in order to feed the CNN. Section \ref{sec:results} describes the tuning of the parameters of the CNN, the accuracy of the algorithm and its performance. The main results are presented in Section \ref{sec:res}, with conclusions drawn in Section \ref{sec:conclusions}.

\section{The {\it COSMODEEP} Convolutional Neural Network}
\label{sec:cnn}
 
Deep Learning builds on the top of neural networks, trying to exploit the inherent structure of the data.  Deep Learning algorithms can take advantage of the spatial correlations of pixels in images, as in the case of Convolutional Neural Networks (CNN), which are among the most successful techniques for image classification. We have adopted the CNN architecture for the implementation of {\cosmodeep}.

\begin{figure*}
\begin{center}
\includegraphics[width=0.99\textwidth]{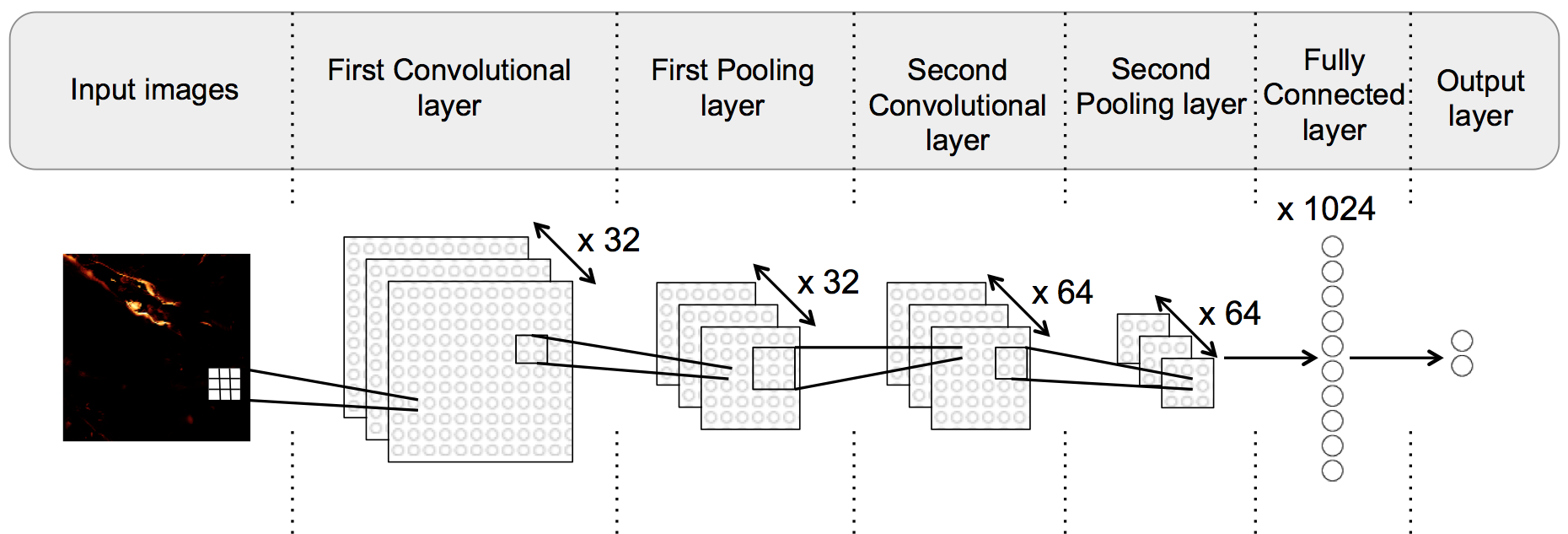}
\caption{The {\cosmodeep} CNN architecture, accounting from one input, five hidden and one output layers}
 \label{figure:cnn}
\end{center}
\end{figure*}

A CNN uses three basic ideas: local receptive fields, shared weights, and pooling. These are combined in a multi-layer architecture whose complexity (``depth'') depends on the problem and on the desired accuracy. The first and the last layers are called {\it input} and {\it output layers}. All the others are called {\it hidden layers}. The CNN network designed for {\cosmodeep} is shown in Figure \ref{figure:cnn}. 

Once input images are loaded in the input layer, each of them is scanned using a local receptive field, which is a small window (e.g. $3 \times 3$ or $5 \times 5$ pixels, in this paper the former is used) moving across all pixels in the image and calculating the {\it activation function}. In the case of our algorithm, this is a {\it ReLU} (Rectified Linear Units) function, which is one of the most successful choices of activation functions for Deep Learning (although several others are possible). It is defined as:
\begin{equation}
\sigma_{i,j} = {\rm max} \bigg( 0,\  b_{i,j} + \sum_{l=1}^M \sum_{m=1}^M w_{l,m} a_{i+l-h,j+m-h} \bigg)
\label{eq:relu}
\end{equation}
where $M$ is the size of the window, $h = (M-1)/2$, $a_{i,j}$ are the pixels, $w_{l,m}$ are the {\it weights} of the network and $b_{i,j}$ are the biases. For every pixel in the image we have the same set of shared $M\times M$ weights plus one additional shared bias. Weights and biases are randomly initialized. The resulting $\sigma_{i,j}$ compose the so-called {\it feature map} at the first hidden layer. Multiple feature maps can be calculated starting from different random initializations of the weights. This leads to a so-called {\it convolutional layer}. Convolutional layers are intended to identify the main features of objects contained in the image, and are usually followed by {\it pooling layers}. Pooling layers take each feature map output from the convolutional layer and calculate a new condensed feature map. It is common practice to use {\it max pooling} or {\it average pooling}, returning the maximum or the average value in a 2$\times$2 input region. The resulting map has half size in each dimension. Pooling is separately applied to each single feature map. It is intended to get rid of the exact positional information of the identified features, focusing on the feature itself, wherever it is placed in the image. 

Convolution and pooling are repeated taking the pooled feature maps at layer N-1 as an input, and producing a new lower resolution set of feature maps at layer N. The information extracted from the images is progressively refined until the final hidden layer. This is usually a fully connected layer that combines and correlates the information refined in the previous layers. At the end, the output layer produces the final answer, which is compared to the correct answer known a priori. Correct answers are part of an image set classified through a labeling procedure performed independently from the CNN (our labeling methodology is described in Section \ref{sec:labeling}). The comparison allows estimating the error through a {\it cost function}. This error is minimized through an optimization process called  {\it training} of the network. Optimization is achieved by calculating corrections to the weights moving along the gradient of the cost function, down toward a minimum value. Such an approach is called {\it gradient descent}. Once corrections have been calculated, they are back-propagated to all the layers of the network, correcting the weights up to the first hidden layer. Back propagation is not performed after each single image, but after a randomly selected sub-set of N training images has been processed and corresponding corrections accumulated. This sub-set of N images is called a {\it mini-batch}, and an optimal setting of N can accelerate the convergence of the algorithm toward the minimization of the error. 

Gradient descent is an iterative process encompassing all the possible mini-batches in the dataset. At each iteration the estimated corrections are weighted by the {\it learning rate} parameter. The learning rate controls how much the weights of the network are adjusted for each mini-batch, influencing the convergence and the accuracy of the algorithm. Small values of the learning rate tend to give more accurate results but lead to slow convergence. Excessively large values may lead to inaccurate results or even divergence.

In order to improve the training, the full training dataset can be used many times. A single pass through the entire dataset is called an {\it epoch}. Each single image is processed by the CNN a number of times equal to the number of epochs during the training, as part of different mini-batches. The optimal number of epochs has to be sufficiently large to extract all the information from the training set, but not too large to slow down the training process or to lead to overfitting (i.e. the CNN starts ``learning'' even from the noise).

A successful Deep Learning network design results from an appropriate combination of the various layers. {\cosmodeep} implements the CNN model shown in Figure \ref{figure:cnn}, consisting of 5 hidden layers, two convolutional layers with 32 and 64 features maps respectively, two pooling layers adopting a max pool algorithm and a fully connected 1024 neurons layer. This rather simple model, accounting for about 700,000 parameters (weights plus biases, their number being independent from the size of the input images) is effective for our purpose. 
The software has been developed using the TensorFlow toolkit  \citep{tensorflow2015-whitepaper} (version currently used: 1.2.1), providing the basic CNN building blocks. TensorFlow deploys a Python API, which have been used for fast and effective prototyping, while the library functions are developed using the C++ programming language for performance purposes. The library efficiently exploits GPUs and provides a distributed interface supporting multi-CPU architectures. 

\section{The Image Set}
\label{sec:training}

Data represents the ``fuel'' of any Deep Learning engine. The availability of a sufficiently {\it large}, {\it labeled} training dataset is one of the most critical aspects in the adoption of a Deep Learning based approach. In the case of data coming from radio observation, sufficiently big datasets are not available, hence we need to generate training  data from scratch exploiting the results of numerical simulations (usually training requires thousands or tens of thousands of images). These results are processed in order to calculate emission at the wavelengths of interest. They are then projected in order to get two-dimensional sky views and further combined with noise and artifacts to mimic actual observations. Finally, the resulting images are automatically labeled. The full procedure is described in the following sections.

\subsection{Images Generation}
\label{sec:generation}

\begin{figure*}
\begin{center}
\includegraphics[width=0.95\textwidth]{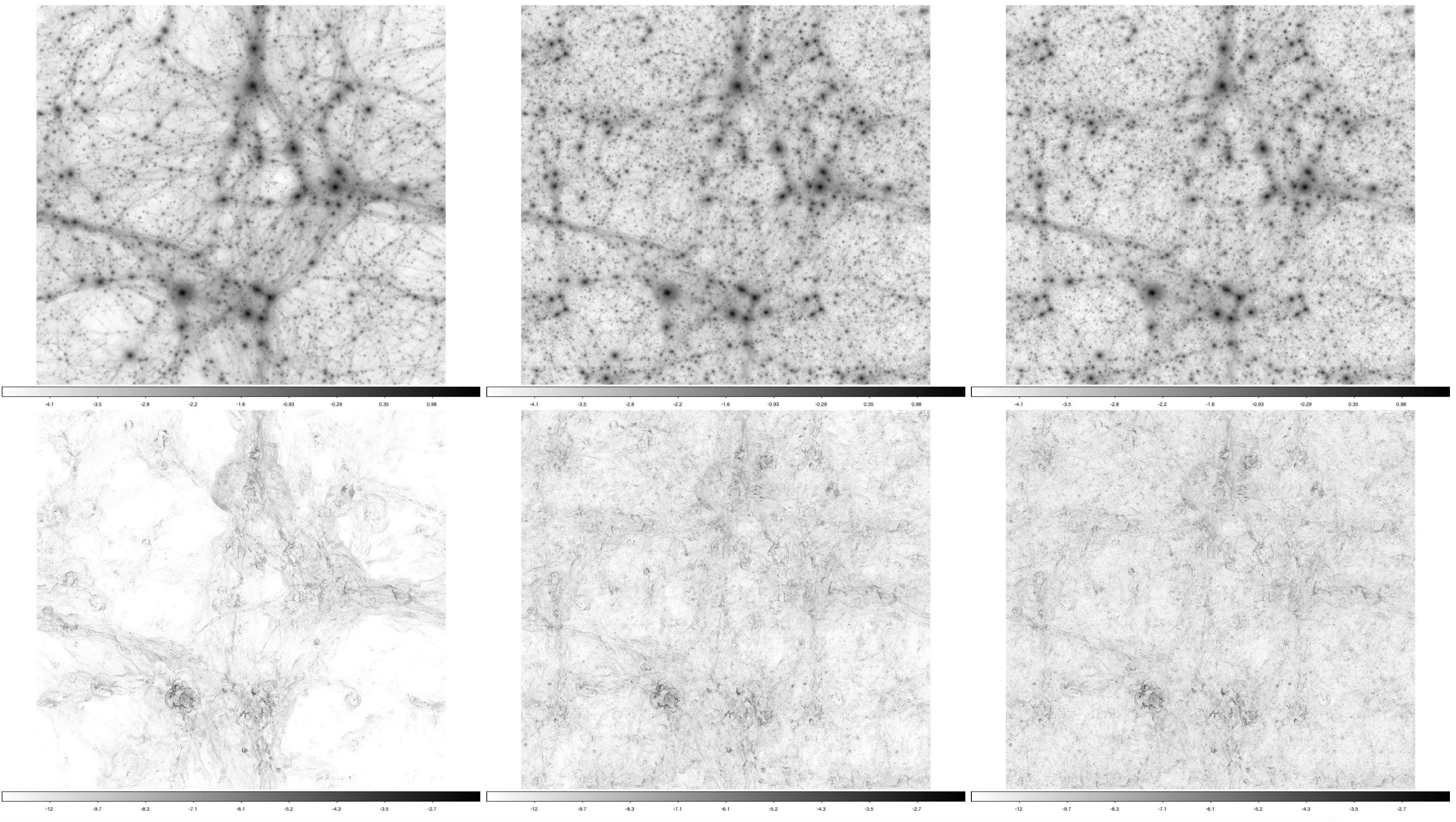}
\caption{Progression of our sky model for a $16 \times 16$ degree area, as a function of the maximum redshift of integration: $z=0.04$ (first panel), $z=0.2$ (second) and $z=0.5$ (third). The top row shows the projected gas density while the second row show the total radio emission, at the frequency of ASKAP.}
 \label{fig:cone}
\end{center}
\end{figure*}

The images created for training need to have size and complexity similar to those expected from real radio observations.  As a test case we considered here the case of a survey made with the Australian telescope ASKAP, the pathfinder of the Square Kilometer Array{\footnote{https://www.atnf.csiro.au/projects/askap/index.html}}. 
ASKAP consists of 36 antennas, each 12m in diameter, with a typical observing frequency of $1.4$ GHz, wide field-of-view, large spectral bandwidth, extremely fast survey speed, and excellent u-v coverage \citep[][]{2008ExA....22..151J}. 
First scientific results obtained with the "BETA" ASKAP configuration based on 6 antennas have already been presented  \citep[][]{2015MNRAS.452.2680S,2016MNRAS.457.4160H}.

We used  as a reference a suite of large cosmological simulations of extragalactic magnetic fields, obtained using the  cosmological code {\it Enzo} \citep[][]{enzo14} as in \cite{va14mhd}, \cite{gh16} and \cite{va17cqg}. Our  simulations evolved a uniform primordial seed field of $B_0=1 ~\rm nG$ (comoving) from high redshift ($z_{\rm in} \geq 40$, the specific figure is dependent on the simulation) in different physical volumes of $200^3$, $100^3$ and $50^3$ Mpc$^3$, in each case with a total number of cells and dark matter particles of $2400^3$.  This model of extragalactic magnetic fields is on the optimistic side, as the assumed initial seed field is at the level of existing upper limits of primordial fields derived from the analysis of the Cosmic Microwave Background \citep[][]{PLANCK2015},  and based on our previous studies on the subject it yields a non-negligible chance of detecting the tip of the iceberg of the magnetic cosmic web \citep[][]{va15radio,va17cqg}. \\

To compute the level of radio emission from cosmic shocks at each redshift, we  assume that shocks can accelerate relativistic particles producing continuum and polarized radio emission \citep[e.g.][]{2011JApA...32..577B}.  We rely on the synchrotron emission model by \cite{hb07}, which requires  the jump condition of each cell undergoing shocks (computed from the simulation), the local value of the magnetic field and the electron acceleration efficiency as a function of Mach number (which is calibrated on shocks internal to galaxy clusters, as in \cite{va15radio}). 

The cosmological model adopted in our simulations has the following parameters: $\Omega_\Lambda=0.692$, $\Omega_M=0.308$, $\Omega_b=0.0478$, $H=67.8 ~\rm km/s$ and $\sigma_8=0.815$. The volumes are resolved with different cell sizes ($83.3$, $41.65$ and $20.82$ kpc, respectively), which is motivated by the fact that our final mock observation is obtained by stacking together the different volumes along the line of sight (with the larger volumes/lower resolution runs being placed at larger distance), which approximately yields a constant angular resolution for all simulations at the corresponding redshifts ($\sim 25-35"$). While this is the intrinsic angular resolution of our simulation (given the starting redshift of the cone integration), we further resampled our images down to a $\approx 10"$ angular resolution for the full ASKAP array configuration.

A detailed description of the procedure adopted to generate mock radio lightcones is given in \cite{va15radio}. To briefly summarise, we create long rectangular volumes covering  $16^{\circ} \times 16^{\circ}$ in the sky, i.e. of the order of $9$ independent ASKAP fields of view. 
Based on \cite{va15radio}, we do not expect to detect a significant amount of radio emission from the cosmic web beyond $z \geq 0.5$, hence we limit our analysis to the the cosmic volume in the range $0.04 \leq z \leq 0.5$
{\footnote{We notice that our lower limit on the integration redshift $0.04$ is motivated because even if there surely are cosmic structures between us $z=0.04$, at the frequency of ASKAP this part of the diffuse emission from the shocked cosmic web gets mostly filtered because of the missing baselines. Due to the vastly larger dataset we need to analyse here, in this work we do not explicitly perform the removal of missing baselines from our mock observations, unlike in previous work \citep[][]{va15radio}. Therefore, we limit by construction our analysis to structures that are located at a large enough redshift  to be properly sampled by ASKAP. For simplicity, we also do not consider the radio artifacts that typically arise as a result of the ``cleaning" procedure of real images \citep[e.g.][]{2014MNRAS.439.4030G}. }} (corresponding to a comoving radial distance of $\approx 1.892$ Gpc). 
This volume is assembled by stacking many simulated boxes along the line of sight, starting with a few replicas of our most resolved $(50  ~\rm Mpc)^3$ box, and then adding several replicas of the  $(100  ~\rm Mpc)^3$ and of the $(200~ \rm Mpc)^3$ volumes.  We first compute the radio emission in the comoving reference frame of each box, and then apply redshift corrections (e.g. cosmological dimming as a function of redshift), assuming the redshift corresponding to the box center for each box. Building the redshift cone, the projected pixel size is adjusted with a cubic interpolation, while the presence of artifacts due to the periodicity of structures along the line of sight is minimised by applying random rotations to each box.  An example of our final result is shown in Figure \ref{fig:cone}. 

Massive halos ($\geq 10^{12} M_{\cdot}$) can be identified by 
running a spherical-overdensity based halo finder at each different redshift, allowing to disentangle the fraction of the radio emission coming from the cosmic web from that coming from galaxy clusters. More details on the procedure to create the mock images can be found in \cite{va15radio}.

We generated a final set of 1000 independent sky model images by applying random rotations to each of the different redshift slices used to produce our lightcones. The maximum spatial resolution achieved in our most resolved box ($\approx 20 ~\rm kpc$) corresponds to an angular resolution of $\approx 0.8 $ kpc at $z=0.04$, i.e. $25"$ per pixel. In order to match the ASKAP angular resolution ($10"$), sky models have been re-sampled to a 2.5 higher resolution. Furthermore, realistic noise has been added at the scale of $\sim 3"$, in order to adequately sample the FWHM of the restoring beam, further re-sampling the sky model at $\sim 8.3$ times higher resolution and then convolving it with a Gaussian profile with a FWHM of $10"$ to emulate a {\it cleaned} ASKAP observation. The assumed noise level is chosen so that $\sigma_{\rm rms}=\sigma_{\rm ASKAP}\approx 10 \rm ~\mu Jy/beam=0.88 \cdot 10^{-7} ~Jy/arcsec^2$. \\

In summary, the described procedure was used to generate $6000 \times 6000$ pixel images, from which smaller $2000 \times 2000$ pixel subsets (corresponding to an ASKAP field of view) have been extracted. These images are indicated as {\it Sky images} and represent the dataset used for the labeling procedure described in Section \ref{sec:labeling}. Adding the noise to the Sky images, the cleaned images are obtained. This data is indicated as {\it Noise images}. The Noise dataset is used for the training of the CNN and to test its performance. 

\subsection{Tiling}
\label{sec:tiling}

\begin{figure*}
\begin{center}
\includegraphics[width=0.75\textwidth]{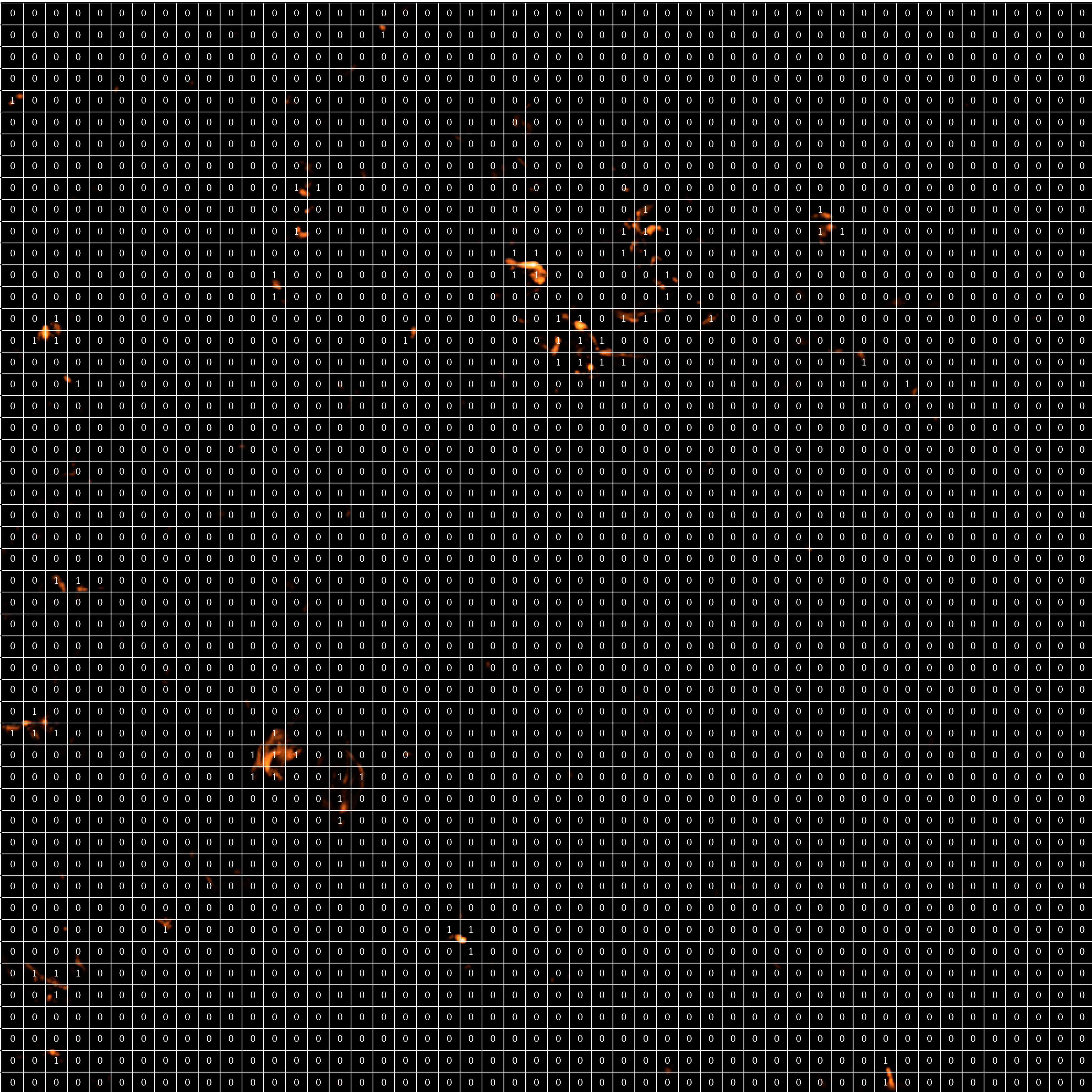}
\caption{A full 2000x2000 pixel Sky image tiled and labeled: label 0 corresponds to a tile without signal (according to our criteria), while label 1 corresponds to a tile with some signal.}
 \label{fig:tiles}
\end{center}
\end{figure*}

The $2000 \times 2000$ pixel images comprise a wide field of view, with tens or even hundreds of potentially interesting objects to detect. In order to identify each single object and its position in the image we have implemented a tiling based procedure that divides each image in small square tiles. The tiles become the actual training dataset of the CNN and each single tile is classified as containing some signal or not. The mosaic of the tiles with signal defines the positions in the sky to observe for radio emitting objects. In order to have a precise localization of the objects, the tile size has to be the smallest possible. However, tiles cannot be so small that objects can not be identified along with their shape and geometric information (i.e. objects should not fill the entire tile). After some experimentation, effective linear tile sizes resulted to be between 40 and 80 pixels. The smallest size, 40 pixels, has been adopted in order to get the highest spatial precision. The result of the tiling procedure is shown in Figure \ref{fig:tiles}, where the mesh composed by the tiles is overlaid on one of the Sky images.

\subsection{Labeling}
\label{sec:labeling}

Labeling is the process of classifying the content of an image so that it can be used to train the Deep Learning model. The labels are assumed to provide the correct values and are used to validate results of the CNN analysis. In our case, images are divided into tiles and each tile has to be defined as containing some radio signal or not. Labeling has, of course, to be performed independently from the Deep Learning network we are training. It is common practice to perform labeling by means of human classification, or using so-called ``bootstrap'' procedures which are semi-automated and still require human supervision. This can be an overwhelming task (especially when hundreds of thousands of images have to be classified), prone to errors and subjectivity, in particular when the target is not a well defined object and noise can blur the content of the image. 

In order to properly label our radio catalog we exploited the Sky images, which are free from noise contamination. A tile containing meaningful signals is positively labeled if the number of pixels emitting above a given flux threshold, $F_{th}$, is larger than $N_{pix}$. We set  $F_{th} = \alpha ~10^{-7} ~\rm Jy/arcsec^2$. For $\alpha=1$, $F_{th}$ is of the order of the 
typical signal to noise ratio considering the (conservative) expected thermal noise level of continuum ASKAP observations for a $10"$ beam (see Section \ref{sec:generation}). Reducing the value of $\alpha$ below 1, we include increasingly fainter sources in the analysis. The parameter $N_{pix}$ sets the minimum size of an object to be labeled as a source, this allows tuning the training of the CNN to identify signals of that size or bigger and excluding point-like sources or objects too small to have meaningful geometric information for the CNN to work with. 

\begin{figure}
\begin{center}
\includegraphics[width=0.5\textwidth]{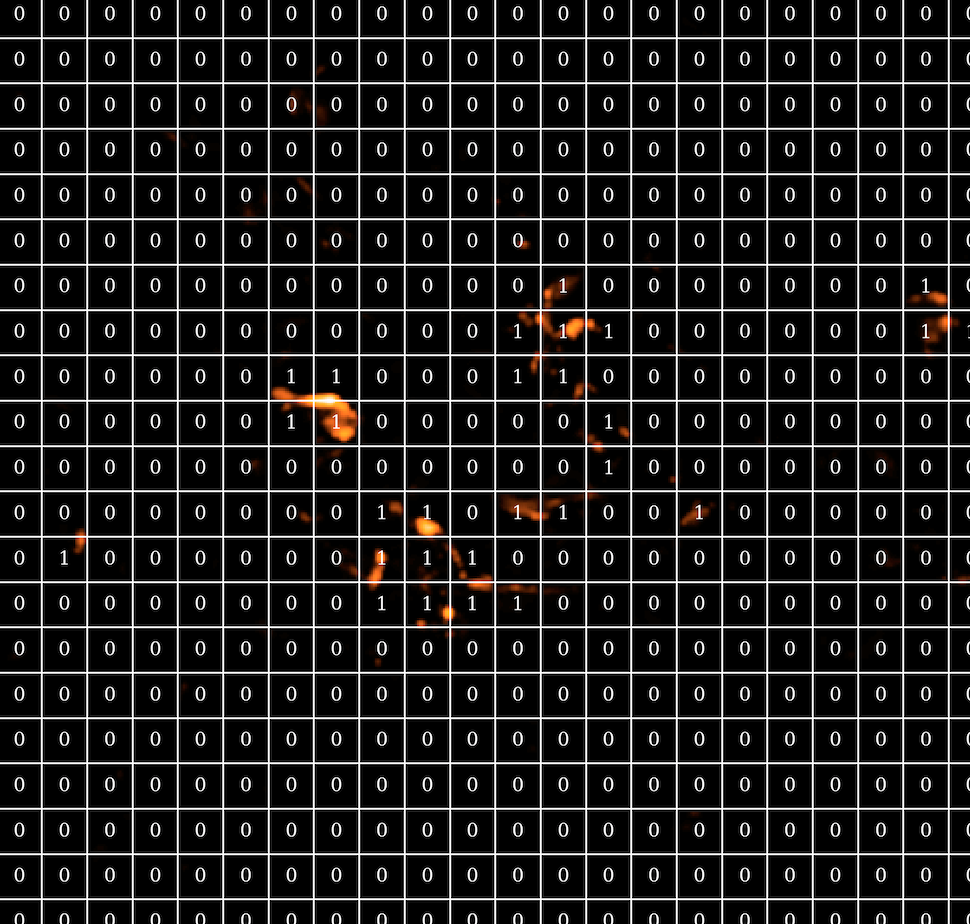}
\caption{A zoom into a $2000 \times 2000$ tiled and labeled Sky image. Label 0 corresponds to a tile without signal, while label 1 corresponds to a tile with some signal.}
 \label{fig:labels}
\end{center}
\end{figure}

The result of the labeling procedure with $\alpha = 1$ and $N_{pix}$= 40 is presented in Figure \ref{fig:labels}, where we show a zoom into one of the $2000 \times 2000$ Sky images. Tiles labeled as "0" have no signal, while those labeled as "1" contain radio sources. 

Once the parameters $F_{th}$ and $N_{pix}$ are set, the labeling process is completely automated. The labels are then used in the training and testing phases where tiles are extracted from the Noise dataset, classified by the CNN, and the results compared to the corresponding labels. 

\section{Parameters Tuning and Performance}
\label{sec:results}

The accuracy of our CNN model has to be properly estimated in order to avoid misinterpretation of the results and incorrect conclusions. Accuracy, in the simplest case, can be defined as the ratio between the number of images correctly classified and the total number of images used in the test. In our case, such estimate is misleading since the number of tiles with no signal can be one or even two orders of magnitude bigger than that of tiles with signal (see Figure \ref{fig:tiles} or \ref{fig:labels}). This definition of accuracy means that simply classifying tiles with no signal would give a very high accuracy regardless of how the tiles with signal are classified. Therefore, we have defined the following accuracy metrics:
\begin{equation}
A_s = N_{sc} / (N_s + N_{vw}),\quad
W_v = N_{vc} / N_v,\quad
W_s = N_{sc} / N_s,
\end{equation}
where $A_s$ gives the fraction of correctly classified signals ($N_{sc}$) over the sum of the total number of tiles that have signal ($N_s$) plus the number of tiles classified as signal but actually are noise ($N_{vw}$). This gives the probability that a tile classified as signal is an actual signal. In other words, it gives the probability that a real signal will be detected by pointing the radio telescope to the region of sky contained by a tile classified as signal. The parameters $W_v$ and $W_s$ measure the relative accuracy for each of the two classes, i.e. the ratio between the number of correctly classified tiles in a class and the total number of tiles belonging to that class. The parameter $N_{vc}$ is the number of tiles with no signal that are correctly classified. Overall, the three parameters describe the accuracy of the CNN, and ideally $A_s = 1$, $W_v = 1$, $W_s = 1$.

\begin{figure}
\begin{center}
\includegraphics[width=0.5\textwidth]{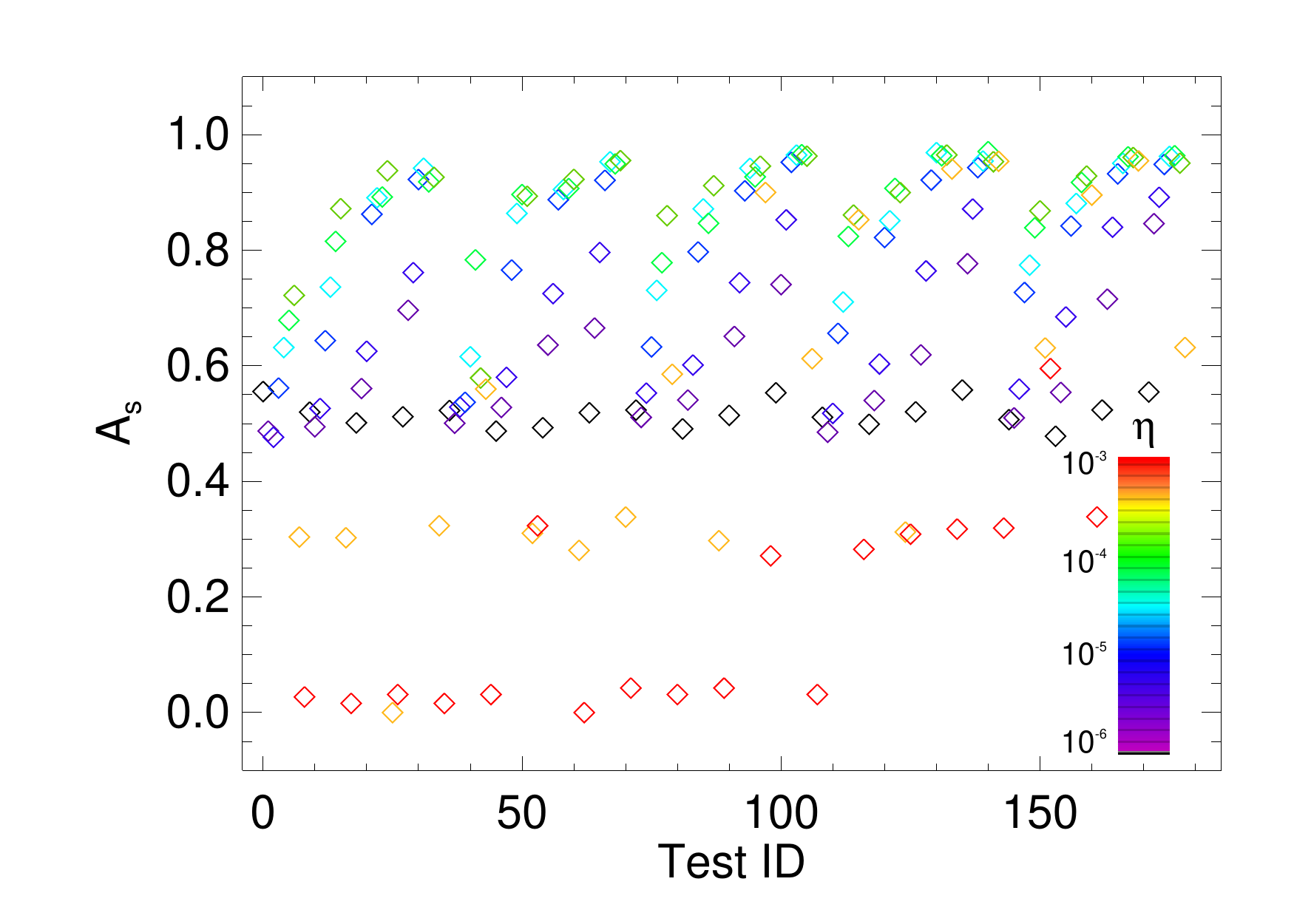}
\includegraphics[width=0.5\textwidth]{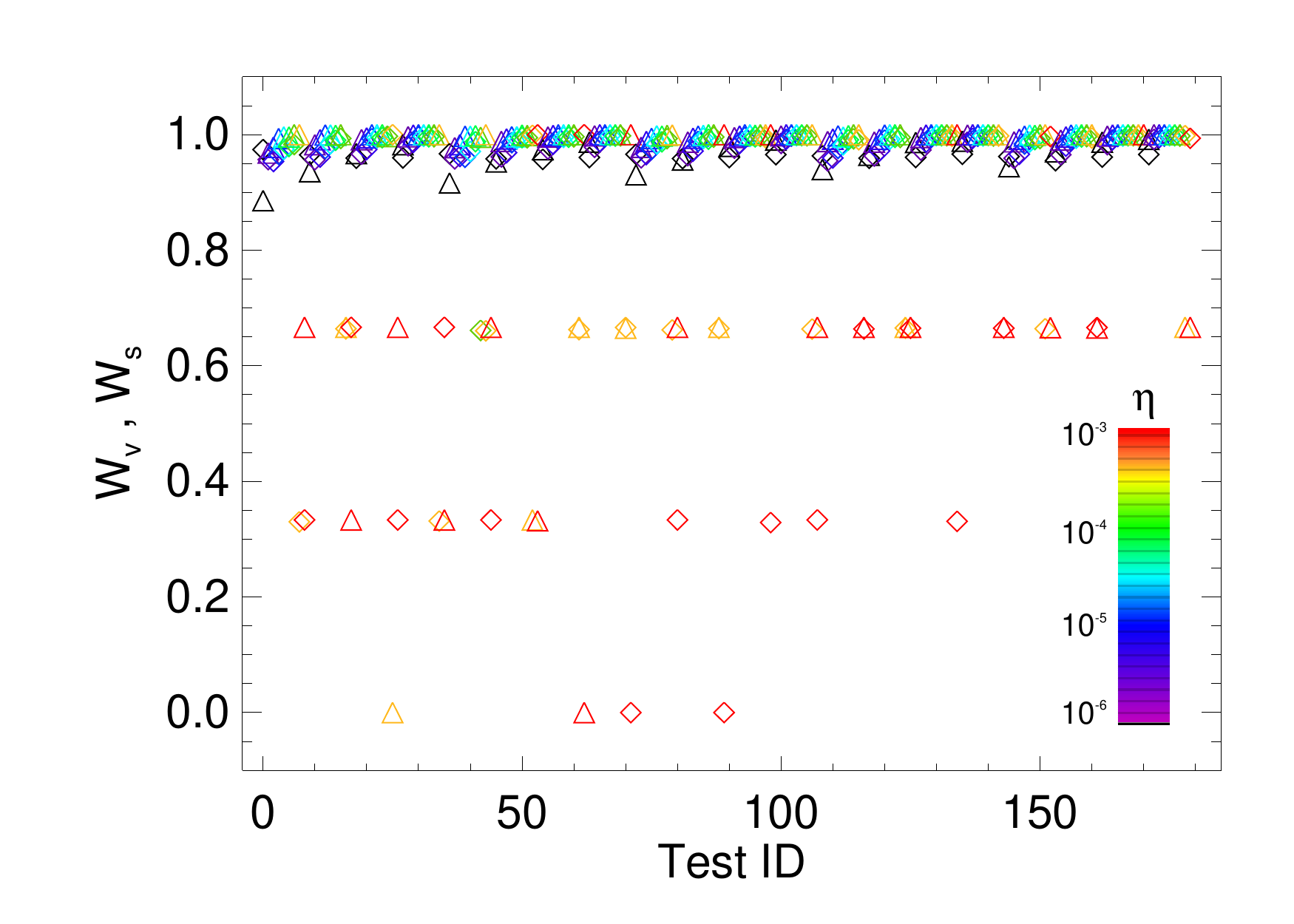}
\caption{Accuracy parameters $A_s$ (top panel), $W_v$ and $W_s$ (diamonds and triangles respectively, bottom panel) in tests with different combinations of learning rate, batch size and number of epochs ($\eta$, $N_b$, $N_e$). The three parameter can take the following values: $\eta$ = [$10^{-6},\ 5\times 10^{-6},\ 10^{-5},\ 3\times 10^{-5},\ 5\times 10^{-5},\ 7\times 10^{-5},\ 10^{-4},\ 5\times 10^{-4},\ 10^{-3}$], $N_b$ = [$15, 30,\ 50,\ 64,\ 100$], $N_e$ = [$10,\ 20,\ 50,\ 100$]. Each combination ($\eta$, $N_b$, $N_e$) is characterized by a different TestID (an integer number between 1 and 180). Colors (in logarithmic scale) represent the parameter $\eta$.}
 \label{fig:test1}
\end{center}
\end{figure}

We have run a number of tests investigating the influence of several parameters on the training of the model, namely the learning rate ($\eta$), the batch size ($N_b$) and the number epochs ($N_e$), all defined in Section \ref{sec:cnn}. For the labeling we have set the parameter $\alpha$, introduced in Section \ref{sec:labeling}, equal to 1. This means that the minimum signal we consider is at the same level of the noise. Furthermore, two different values for $N_{pix}$ have been tested: $N_{pix1}=40$, which focuses on extended sources, and $N_{pix2}=9$, which corresponds to the size of the local receptive field of the CNN, setting the resolution of the method. 

All the tests have been performed on an Intel Xeon E5-2690 ``Haswell'' CPU running at 2.60GHz (12 cores, 64GB RAM) equipped with NVIDIA Tesla P100 with 16GB HBM2 memory, which is effectively exploited by {\cosmodeep} through TensorFlow. The computing environment is part of the Piz Daint supercomputer, available at the Swiss National Supercomputing Center in Lugano (operated by ETH Zurich). The size of the input dataset is about 6.5 GB, mostly used as training set and a small fraction dedicated to testing and validation. 

The following workflow is implemented for the training and testing of the CNN:
\begin{enumerate}
\item Sky and Noise images are read from files stored on disk;
\item negative pixels are set to a small floor value (typically $10^{-11}$ Jy/arcsec$^2$), and the logarithm of each pixel is calculated in order to reduce the dynamical range of the emissivity, which typically spans 10 orders of magnitude ($10^{-11}-10^{-2}$ Jy/arcsec$^2$), avoiding issues related to floating point precision; 
\item the results are normalized so that each image has values between 0 and 1;
\item images are divided into tiles;
\item using the Sky tiles, each tile is labeled according to the procedure described in Section \ref{sec:labeling};
\item tiles are serialized to feed the CNN;
\item tiles are offloaded to the GPU (in chunks, in order to avoid GPU memory overflows) and there processed by the CNN for the training; 
\item the trained network is finally tested and its accuracy calculated.
\end{enumerate}

The resulting accuracy for the case $N_{pix1}$ is presented in Figure \ref{fig:test1}.
The top panel shows $A_s$ as obtained for the different combinations of $\eta$, $N_b$ and $N_e$. Colors highlight the dependence on the learning rate.
In a number of cases, the accuracy is above 0.9. The highest values for $A_s$ are obtained by setting the learning rate bigger than $10^{-5}$. However, for $\eta > 5\times 10^{-4}$ accuracy drops and convergence is not reached. For $\eta < 10^{-6}$ the convergence is slow. 
The mini-batch size progressively grows with the TestID, starting from $N_b = 15$ for TestID $< 20$, up to $N_b = 100$ for $160\le$ TestID $< 180$, stepping up every 20 TestIDs. Its influence can be seen in the overall trend of the accuracy to slightly increase when shifting toward higher TestIDs, from left to right (to higher mini-batch sizes). Accuracy is also improved by increasing the number of epochs. We have performed tests using four different numbers of epochs ($N_e$ = 15, 20, 50, 100). The accuracy of the method grows at larger $N_e$, as can be seen from the tendency of the accuracy of matching colored (i.e. $\eta$) point data to have higher $A_s$ moving toward larger values of TestID. This continues until the mini-batch size is updated to a new value, when $A_s$ drops. The same behavior is shown in the bottom panel of the figure for the parameters $W_s$ and $W_v$, although most of data points are close to unity and the trend is less recognizable. 
Tests with the highest values of $\eta$ (yellow and red points) have low accuracy and do not present any trend varying both the number of epochs and the mini-batch size, showing that their accuracy cannot be improved by tuning the two parameters.

The bottom panel of Figure \ref{fig:test1}, which shows the relative accuracy parameters $W_s$ and $W_v$, confirms the results discussed above. In most cases {\cosmodeep} is capable of successfully classifying more than 99\% of both regions with signals and empty regions. 

When decreasing the pixel threshold to $N_{pix2}$ pixels (not shown), objects at the limit of the resolution of the method are included, leading to a slightly lower accuracy. The overall trends, however, are the same as in the $N_{pix1}$ case.

\begin{figure}
\begin{center}
\includegraphics[width=0.5\textwidth]{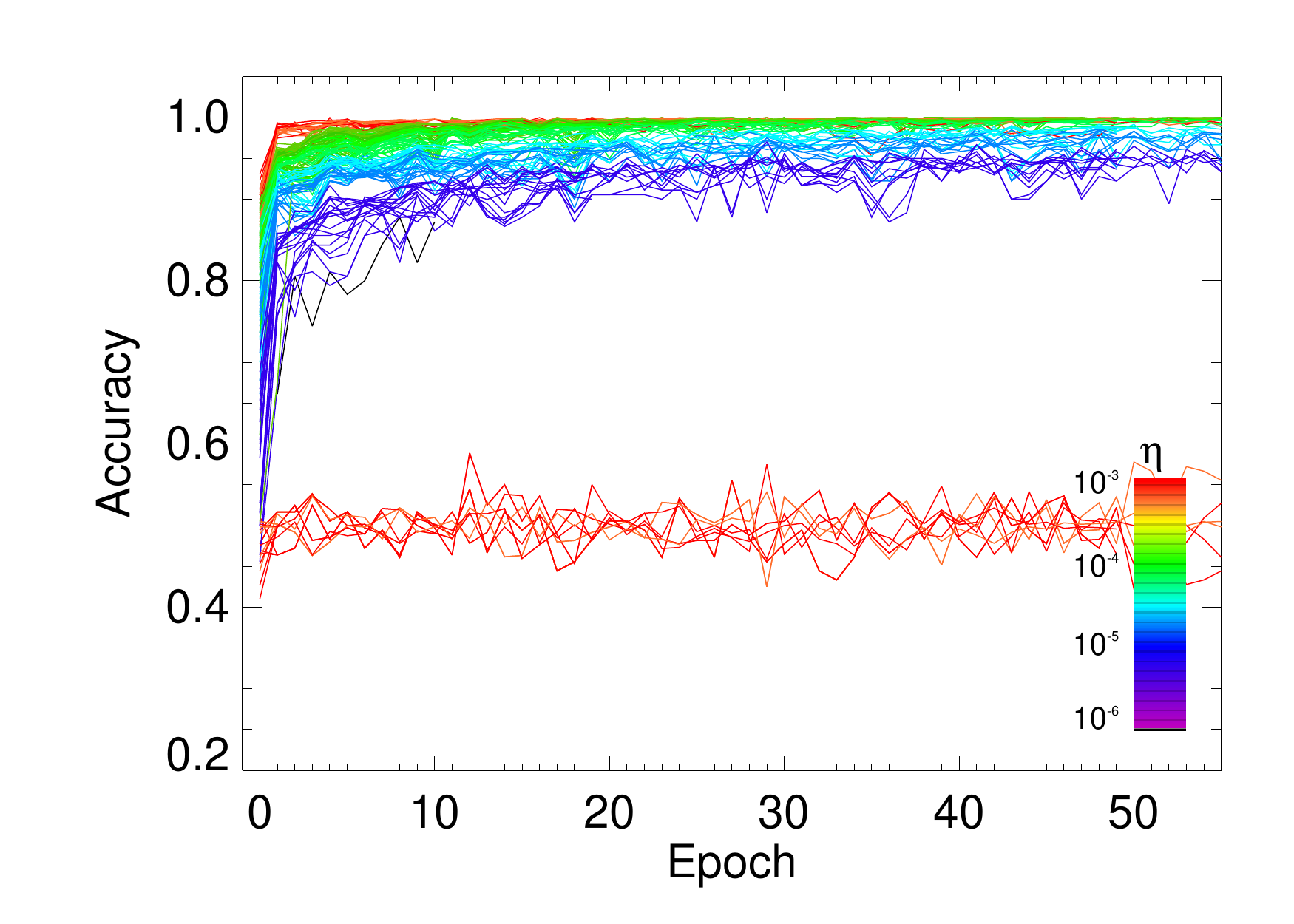}
\caption{Convergence of the training process as a function of the number of epochs, measured by an Accuracy parameter defined as the fractional difference between the tiles correctly classified and the total number of tiles used for measure. A constant value of the Accuracy indicates that the training cannot improve more. The optimal value for the Accuracy is 1, which indicates that all the images are correctly classified. Colors (in logarithmic scale) show the dependency from the learning rate $\eta$.}
 \label{fig:epochs}
\end{center}
\end{figure}

Figure \ref{fig:epochs} shows the convergence of the training process as a function of the epoch, for the case $N_{pix1}$, with different settings of $\eta$ and $N_b$. The set-ups with $5\times 10^{-5} \le \eta \le 10^{-4}$ and $N_b \ge 30$ (green and blue curves in the figure), have the fastest convergence toward an accuracy close to 1. For this specific test the accuracy is calculated as the ratio between the number of tiles correctly classified and the total number of tiles used for the test. For $\eta < 5\times 10^{-5}$ the algorithm converges but very slowly, while for $\eta > 10^{-4}$ most of the tests do not converge, the accuracy fluctuates around 0.5 which corresponds to random classification. Few cases with $\eta \geq 5\times 10^{-4}$ and  $N_b > 50$ converge faster than in all the other cases. However, their $A_s$, $W_s$ and $W_v$ are low, proving that the training is not actually effective, only tiles without signal being correctly classified. Similar results are obtained for $N_{pix2}$.

In terms of computational requirements, the training takes around 1800-2500 seconds to complete, the time depending essentially from the number of epochs. The trained network can be stored in files and reloaded for later usage for image classification. The CNN network load and setup time is independent from the number of images to classify, depending only from the size of the network, while the classification stage scales linearly with the image size. For our $2000 \times 2000$ pixel images, the estimated classification performance is $10.4\pm 0.2$ images/sec.

\section{Results}
\label{sec:res}

\begin{figure*}
\begin{center}
\includegraphics[width=1\textwidth]{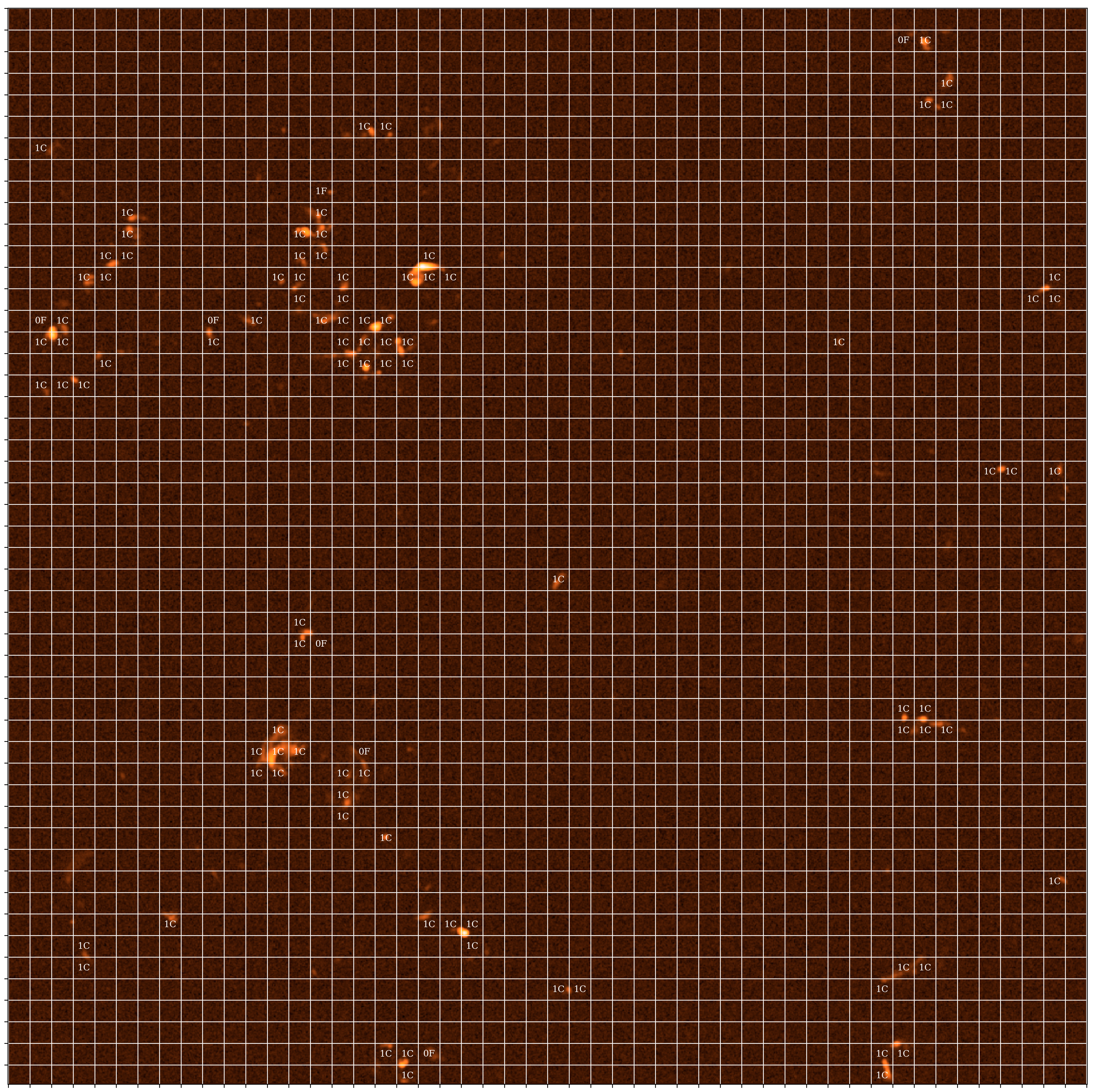}
\caption{A full $ 2000 \times 2000$ pixel image classified by {\cosmodeep}. Label 1C refers to correctly classified signals, 1F indicates tiles with signal incorrectly classified as noise (false negatives), 0F indicates tiles with no signal and wrongly classified (false positives). Unlabeled tiles indicates pure noise tiles correctly classified.}
 \label{fig:fullclass}
\end{center}
\end{figure*}

The effectiveness of {\cosmodeep} in detecting faint, diffused radio sources in noisy images has been analyzed on a subset of images, the test dataset, never used for the training. The CNN has been trained for the two different choices of $N_{pix}$, indicated as $N_{pix1}$ (40 pixels) and $N_{pix2}$ (9 pixels). The parameters listed in Table \ref{tab:params} have been set in order to optimize the performance of the CNN according to the analysis performed in Section \ref{sec:results}. Figure \ref{fig:fullclass} shows one of the test images, with tiles labeled as follows: tiles classified as ``1C" are signals correctly detected by the CNN, ``1F" indicates tiles with signal but classified as pure noise (false negatives), tiles with ``0F" are pure noise tiles classified as signal (false positives). The remaining tiles, labeled as "0C", are correctly classified as pure noise and for clarity their label is not displayed in the image.

In the case $N_{pix1}$, we get the following accuracy estimates: $A_s = 0.9088\pm 0.0090$, $W_s = 0.9827\pm 0.0003$, $W_v = 0.9974\pm 0.0003$. On average our classifier misses around 1 to 2 tiles with signal ($40 \times 40$ pixels) per $2000 \times 2000$ pixel image (out of a total number of 2500 tiles), and it misclassifies around 6 pure noise tiles per image. The CNN proves to be effective in detecting extended objects (at least bigger than 40 pixels) missing less than 2\% of them. The accuracy improves for regions without emission, even if the absolute number of false positives is larger than that of false negatives. The case $N_{pix2}$ returns $A_s = 0.8800\pm 0.0062$, $W_s = 0.9710\pm 0.0062$, $W_s = 0.9957\pm 0.0003$, with, on average, around 2 to 3 tiles misclassified as 1F and around 10 as 0F per image. The total number of tiles with signal increases from around 600 in the $N_{pix1}$ case, to around 800 in the $N_{pix2}$, since smaller objects are classified as sources by the labeling procedure. Such smaller objects are also more challenging to recognize, being at the limit of the resolution of the CNN. This explains the slight decrease of accuracy in the $N_{pix2}$ case. 

\begin{table}
\centering
\begin{tabular}{l|r}
Parameter & Value \\\hline\hline
Image size (pixels) & 2000$\times$2000 \\
Tile size (pixels) & 40$\times$40 \\
$N_{pix1}$ & 40\\
$N_{pix1}$ & 9\\
$F_{th}$ (Jy/arcsec$^2$) & $10^{-7}$\\
$\eta$ & $7\times 10^{-5}$ \\
$N_b$ & 100 \\
$N_e$ & 50 \\
\hline
\end{tabular}
\caption{Set-up of the CNN models.}
\label{tab:params}
\end{table}

\begin{figure*}
\begin{center}
\includegraphics[width=0.25\textwidth]{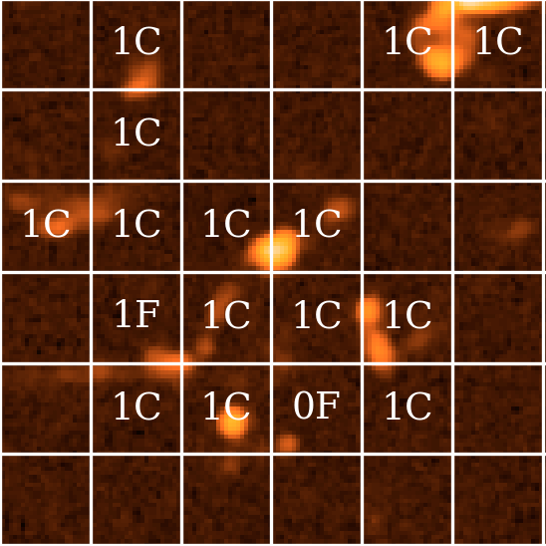}
\includegraphics[width=0.25\textwidth]{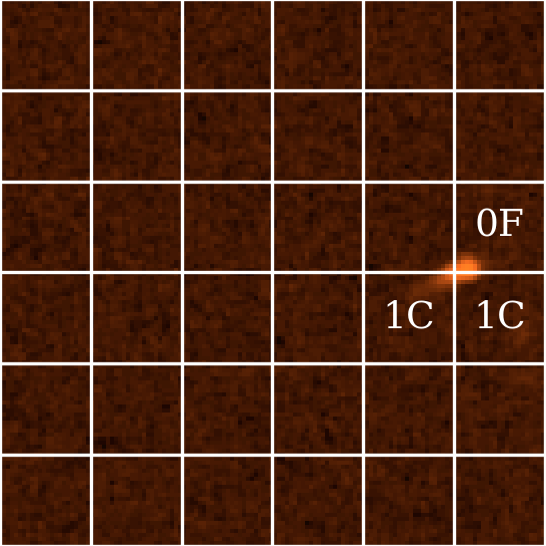}
\includegraphics[width=0.25\textwidth]{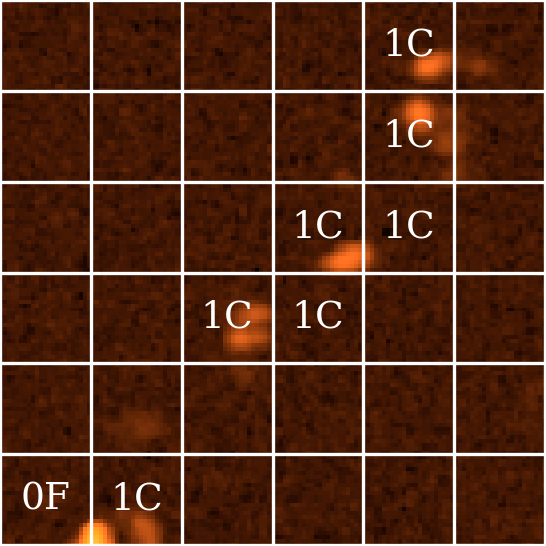}
\includegraphics[width=0.25\textwidth]{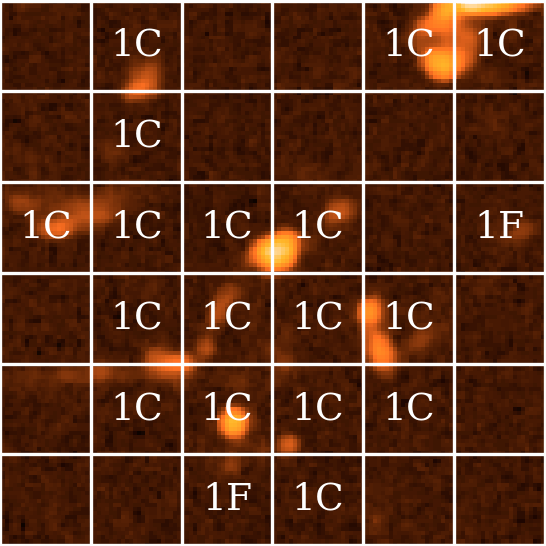}
\includegraphics[width=0.25\textwidth]{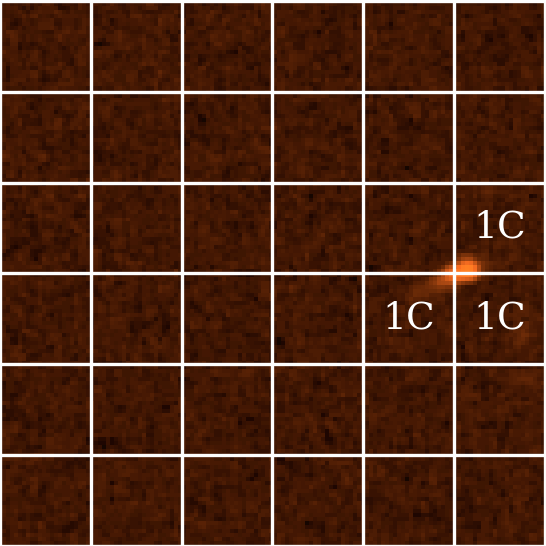}
\includegraphics[width=0.25\textwidth]{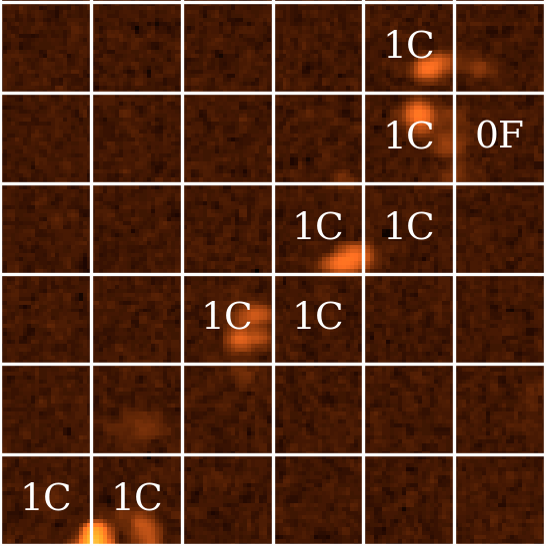}
\caption{Mosaics of tiles extracted from the full images for $F_{th} = 10^{-7}$ (Jy/arcsec$^2$), with different kind of classification. Label 1C refers to correctly classified signals, 1F indicates false negatives, 0F false positives. Unlabeled tiles indicates pure noise tiles correctly classified. Upper row refers to $N_{pix1}$, bottom row to $N_{pix2}$}
 \label{fig:class}
\end{center}
\end{figure*}

Figure \ref{fig:class} zooms into three regions extracted from one test image, in the $N_{pix1}$ (upper row) and in the $N_{pix2}$ (lower row) cases. In the top-left panel, we see a region with a prominent cluster of galaxies with a clear pattern of shock waves moving outwards from the cluster center.  In the $N_{pix1}$ case most of the tiles with signal are correctly classified (1C) and one false positive is present. The false positive (tile labeled as 0F) identifies a tile that is actually part of the cluster, but the number of pixels above $F_{th}$ is less than $N_{pix1}$. Lowering the pixel threshold to $N_{pix2}$, the same tile results to be classified as 1C. Therefore, the false positive in the $N_{pix1}$ case follows from the labeling procedure, and not from the CNN. In the top-central panel, a small object is detected, split into three tiles. In the $N_{pix1}$ case one of the tiles is again misclassified due to the labeling method, showing that the CNN can indeed correctly detect and classify sources below the pixel threshold it has been trained for. The right panels show the example of a filament connected to a galaxy cluster, which appears in the bottom-left corner of the image. The structure, albeit elongated and discontinuous, can be properly identified by the CNN. A false positive is present in the $N_{pix2}$ case and is again a shortcoming of the labeling procedure (i.e. there are not enough pixels above the flux threshold in the tile), and not a error of the CNN. 

In the top-left panel we can also see how in the $N_{pix1}$ case, several tiles contain sources which get labeled as noise, and are not detected by the CNN. A couple of these tiles are classified as false negative in the $N_{pix2}$ case (bottom-left panel). The sources contained in these tiles are larger than $N_{pix1}$ pixels but too faint, hence they are labeled as noise. Accordingly, the CNN is trained to classify those kind of objects as noise. Reducing the pixel threshold to $N_{pix2}$, a sufficiently large number of pixels in both tiles are brighter than $F_{th}$, hence they are labeled as sources. However, the CNN is not able to detect them as 
most of the source still emitting below the flux threshold the CNN has been trained for.
 
In order to investigate the influence of $F_{th}$ in detail, we have repeated the whole training and testing procedure reducing the value of the flux threshold progressively to 0.75, 0.5 and $0.1 \times 10^{-7}$(Jy/arcsec$^2$), trying to detect signals with smaller and smaller signal to noise ($SN$) ratio. All the other parameters of the CNN are unchanged (see Table \ref{tab:params})

The results, presented in Table \ref{tab:flux}, show that for the $N_{pix1}$ model, down to $F_{th} = 0.5\times 10^{-7}$(Jy/arcsec$^2$), $A_s$ is bigger than 0.9 and both the parameters $W_v$ and $W_s$ are close to 1. For $N_{pix2}$, the accuracy parameter $A_s$ is slightly lower, due to the presence of smaller objects to be detected, but still of the order of 0.9. The $W_v$ parameter is very close to unity, while $W_s$ is around 0.97. In both cases the best accuracy is achieved for $F_{th,opt} = 0.75\times 10^{-7}$(Jy/arcsec$^2$), so below the rms noise, with on average around 0.5 and 3 false negatives and 6 and 7 false positives per image for the $N_{pix1}$ and $N_{pix2}$ cases respectively. The higher accuracy reached at $F_{th,opt}$ is due to the inclusion in the training set of faint sources that at higher flux thresholds are labeled as noise but that are detectable by the CNN. The mismatch between labeling and classification leads to a slightly less efficient training with some loss of accuracy at $F_{th} = 10^{-7}$(Jy/arcsec$^2$). Below $F_{th,opt}$ the accuracy decreases due to the presence of smaller and fainter sources, blurred by the noise. At $F_{th} = 0.1\times 10^{-7}$(Jy/arcsec$^2$) accuracy drops, in particular for $N_{pix2}$, with $A_s$ slightly bigger than 0.8, and, on average, around 60 misclassified tiles per image.

\begin{figure*}
\begin{center}
\includegraphics[width=0.25\textwidth]{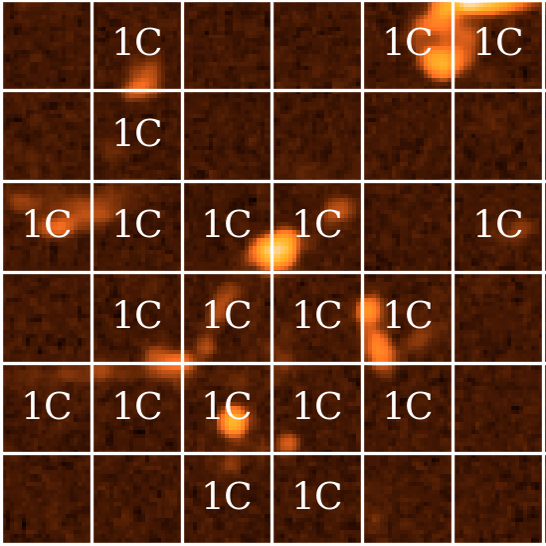}
\includegraphics[width=0.25\textwidth]{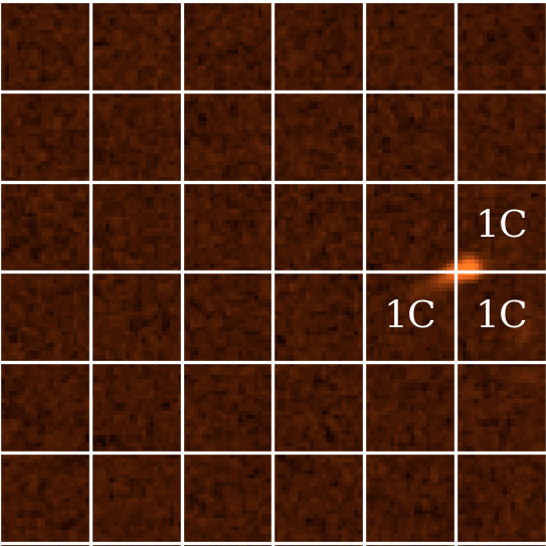}
\includegraphics[width=0.25\textwidth]{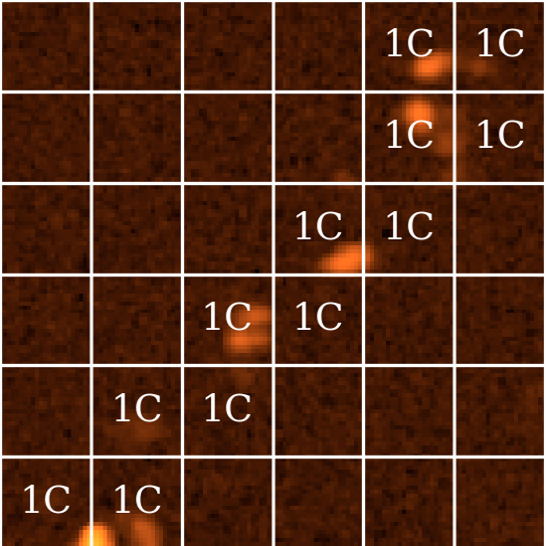}
\includegraphics[width=0.25\textwidth]{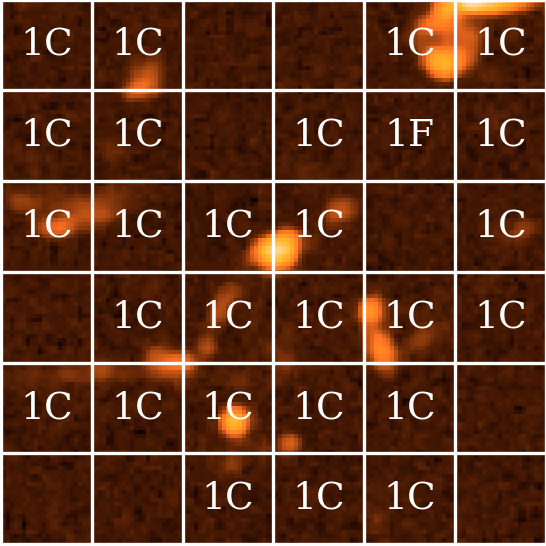}
\includegraphics[width=0.25\textwidth]{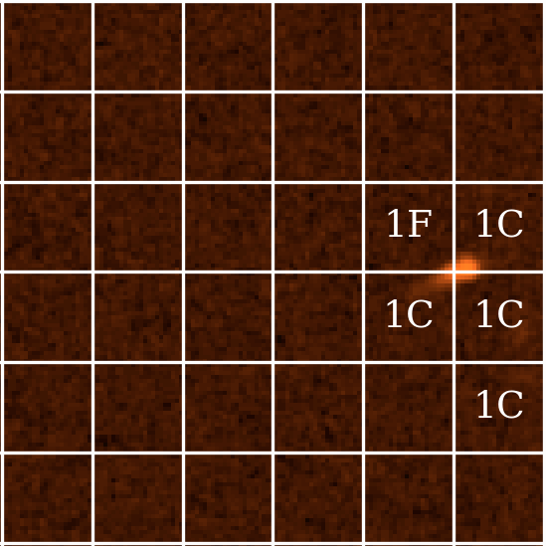}
\includegraphics[width=0.25\textwidth]{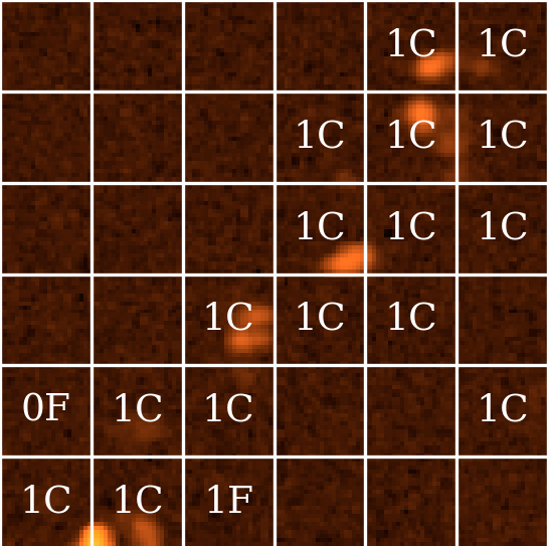}
\caption{Mosaics of tiles extracted from the full images for the case $N_{pix2}$ at different values of $F_{th}$. For the top row $F_{th} = 0.75\times 10^{-7}$ (Jy/arcsec$^2$), while for the bottom row $F_{th} = 10^{-8}$ (Jy/arcsec$^2$)}
 \label{fig:thcomp}
\end{center}
\end{figure*}

Figure \ref{fig:thcomp} shows the same regions of Figure \ref{fig:class} for the case $N_{pix2}$, but at $F_{th,opt}$ (top row) and $F_{th} = 0.1\times 10^{-7}$(Jy/arcsec$^2$). At $F_{th,opt}$ all the tiles are correctly classified. The false negatives on the bottom-left panel of Figure \ref{fig:class} are now correctly classified, since the CNN is trained to recognize those faint sources. At $F_{th} = 0.1\times 10^{-7}$(Jy/arcsec$^2$), more tiles are labeled as containing signal and classified as 1C. As expected, a few incorrect classifications appear due to the presence of extremely faint and small objects the CNN is not able to detect or the labeling schema neglects.

\begin{table*}
\centering
\begin{tabular}{l|l|c|c|c}
Model & $F_{th}$ (Jy/arcsec$^2$) & $A_s $ & $W_s$ & $W_v$  \\\hline\hline
$N_{pix1}$ & $10^{-7}$ & $0.9088\pm 0.0090$ & $0.9827\pm 0.0003$ & $0.9974\pm 0.0003$  \\
& $0.75\times 10^{-7}$ & $0.9232\pm 0.0046$ & $0.9947\pm 0.0002$ & $0.9972\pm 0.0002$  \\
& $0.5\times 10^{-7}$ & $0.9159\pm 0.0070$ & $0.9958\pm 0.0005$ & $0.9960\pm 0.0004$  \\
& $0.1\times 10^{-7}$ & $0.8934\pm 0.0072$ & $0.9521\pm 0.0005$ & $0.9933\pm 0.0005$  \\
\hline
$N_{pix2}$ & $10^{-7}$ & $0.8800\pm 0.0062$ & $0.9710\pm 0.0062$ & $0.9957\pm 0.0003$  \\
& $0.75\times 10^{-7}$ & $0.9175\pm 0.0051$ & $0.9754\pm 0.0037$ & $0.9969\pm 0.0001$  \\
& $0.5\times 10^{-7}$ & $0.9022\pm 0.0046$ & $0.9797\pm 0.0044$ & $0.9948\pm 0.0001$  \\
& $0.1\times 10^{-7}$ & $0.8181\pm 0.0035$ & $0.8848\pm 0.0023$ & $0.9890\pm 0.0011$  \\
\hline
\end{tabular}
\caption{Accuracy parameters of {\cosmodeep} for different values $F_{th}$, corresponding to models $N_{pix1}$ and $N_{pix2}$.}
\label{tab:flux}
\end{table*}

\begin{figure*}
\begin{center}
\includegraphics[width=0.99\textwidth]{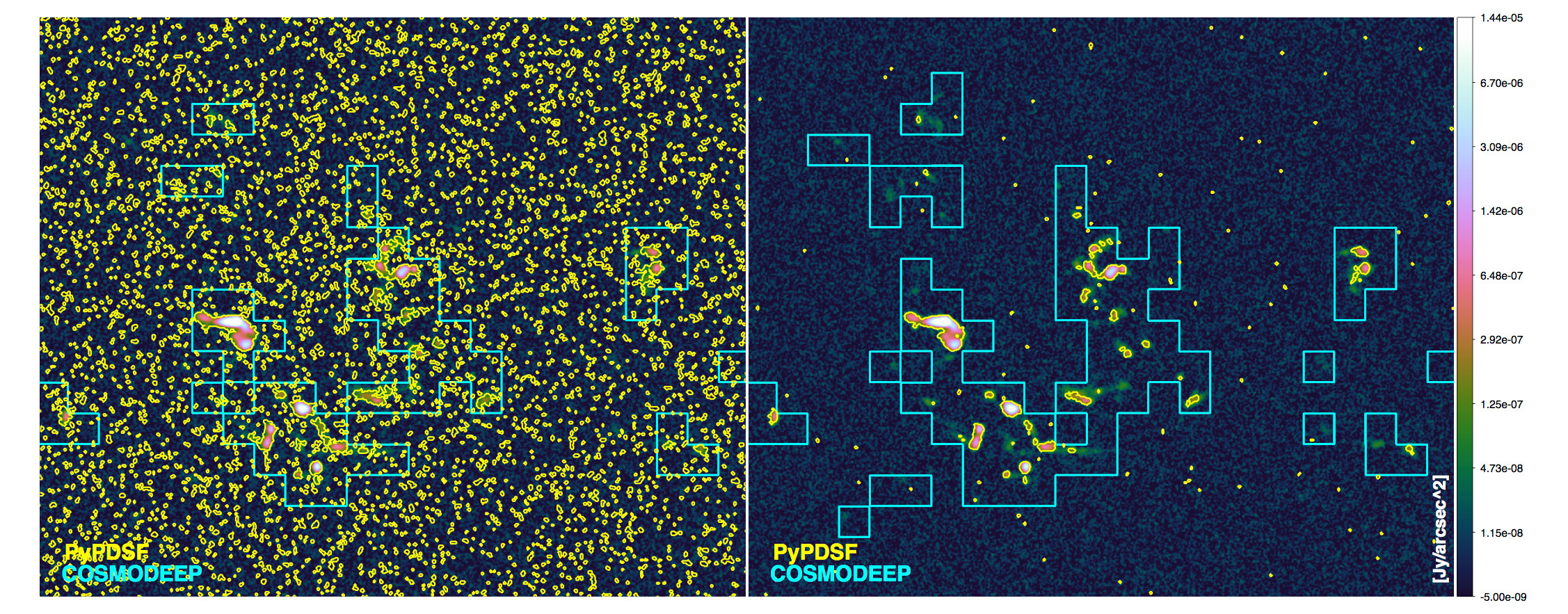}
\includegraphics[width=0.99\textwidth]{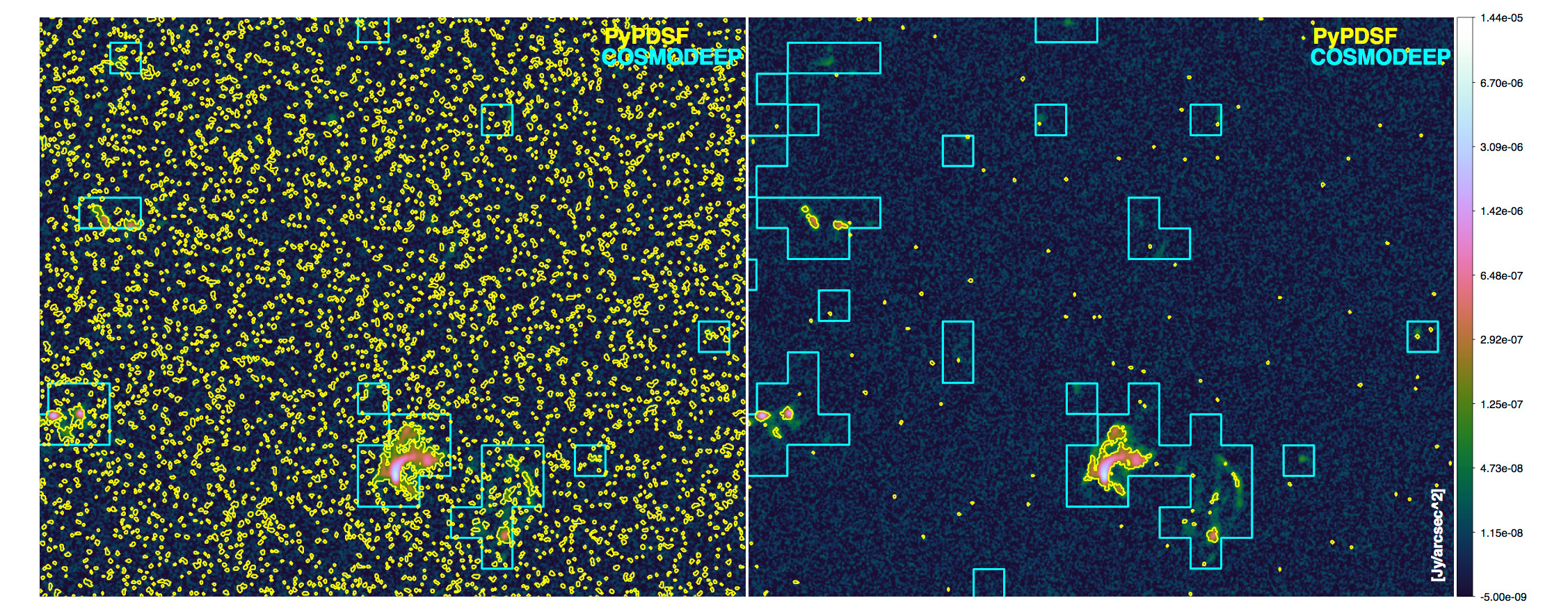}
\caption{Close up view of 2 sample $\sim 2^\circ \times 1.5^\circ$ simulated maps, showing our noise-added mock sky model (colors, in units of $\rm Jy/arcsec^2$). For the same sky model, we show with cyan contours all tiles correctly identified by {\cosmodeep} using either $N_{\rm pix}=40$ and $F_{\rm th}=0.1 \times 10^{-7} \rm ~Jy/arcsec^2$ (left) or $N_{\rm pix}=9$ and $F_{\rm th}=0.5 \times 10^{-7} \rm ~Jy/arcsec^2$ (right). In the same panels, we also show 
with yellow contours the "islands of signal" identified by PyBDSF 
assuming either threshold of $1.5 ~\rm SN$ or 
$3.0 ~\rm SN$. }
 \label{fig:comparison}
\end{center}
\end{figure*}

We compared the results of {\cosmodeep} with those obtained using PyBDSF (the Python Blob Detector and Source Finder, see http://www.astron.nl/citt/pybdsf), which is a python-based tool to decompose radio interferometric images into sources. 
Since PyBDSF is designed to work on real images, several parameters can be set by the users to distinguish e.g. regions in the image with different noise properties (for example, different noise due to imaging artifacts around strong sources). For our purposes, we have run the tool adopting standard input parameters, setting the noise level to a constant value of $\sigma_{\rm rms} = 10 ~\mu \rm Jy/beam$. 
We  considered as islands of signal, regions in the image that show at least 9  contiguous pixels above the assumed flux threshold (to allow a close comparison with {\cosmodeep}), and tested variations in rms noise level from $0.75$ to $3.0$.
Using the above parameters, we identified islands with signals of contiguous emission as shown by the green contours in Fig.\ref{fig:comparison}. In a second step, the algorithm fits a Gaussian profile to each island, in order to further decompose them into shapelets.  The final result is a catalog of  sources with  positions, sizes, and flux densities of each source. However, this second step is not required for our purposes, as the islands already identify the regions in the image where emission above threshold is detected by the algorithm.  

We compare in Fig. \ref{fig:comparison} the results of {\cosmodeep} to those of PyPDSF at different values of $F_{th}$, for two $\sim 2^\circ \times 1.5^\circ$ fields, featuring several diffuse emission patches.  Sources identified by PyBDSF for different choices of $F_{th}$ are given in green contours,  while the rectangular tiles identified by {\cosmodeep} are marked by the white contours. We find a tight correspondence between the islands of signal identified by PyBDSF imposing a threshold of $SN \geq 3.0$ and the results of {\cosmodeep} for $F_{th} \ge 0.5 \times 10^{-7}$Jy/arcsec$^2$ with a lower bound of $9$ pixels for the size of structures. Interestingly, lowering the threshold to $F_{th} \ge 0.1 \times 10^{-7}$Jy/arcsec$^2$ and using $40$ pixels for the size of structures allows {\cosmodeep} to correctly identify a few more fainter low-surface brightness structures in the sky model, while lowering the threshold in  PyBDSF to $SN \geq 1.5$ causes the software to detect a large number of spurious noise fluctuations, randomly spread across the field (see right panels in Fig.~\ref{fig:comparison}).

While further ad-hoc improvement in PyBDSF is surely possible, this test shows that the two algorithms can give consistent results on the high signal-to-noise end of the distribution of sources, while with very little tuning {\cosmodeep} can go significantly below the ``standard" $2-3 ~ \sigma_{\rm rms}$ level for the detection of {\it real} diffuse emission from the cosmic web. On the other hand,  the spatial resolution of {\cosmodeep} is limited to the tile size, which prevent us from exactly describing the shape of these emission regions. Based on the above results, it seems possible to design a combined approach  in future work, where {\cosmodeep} and PyBDSF may be applied to large datasets in two different steps, to better trace the location of diffuse emission structures at a scale comparable to the restoring beam of observations. \\

\begin{figure}
\begin{center}
\includegraphics[width=0.5\textwidth]{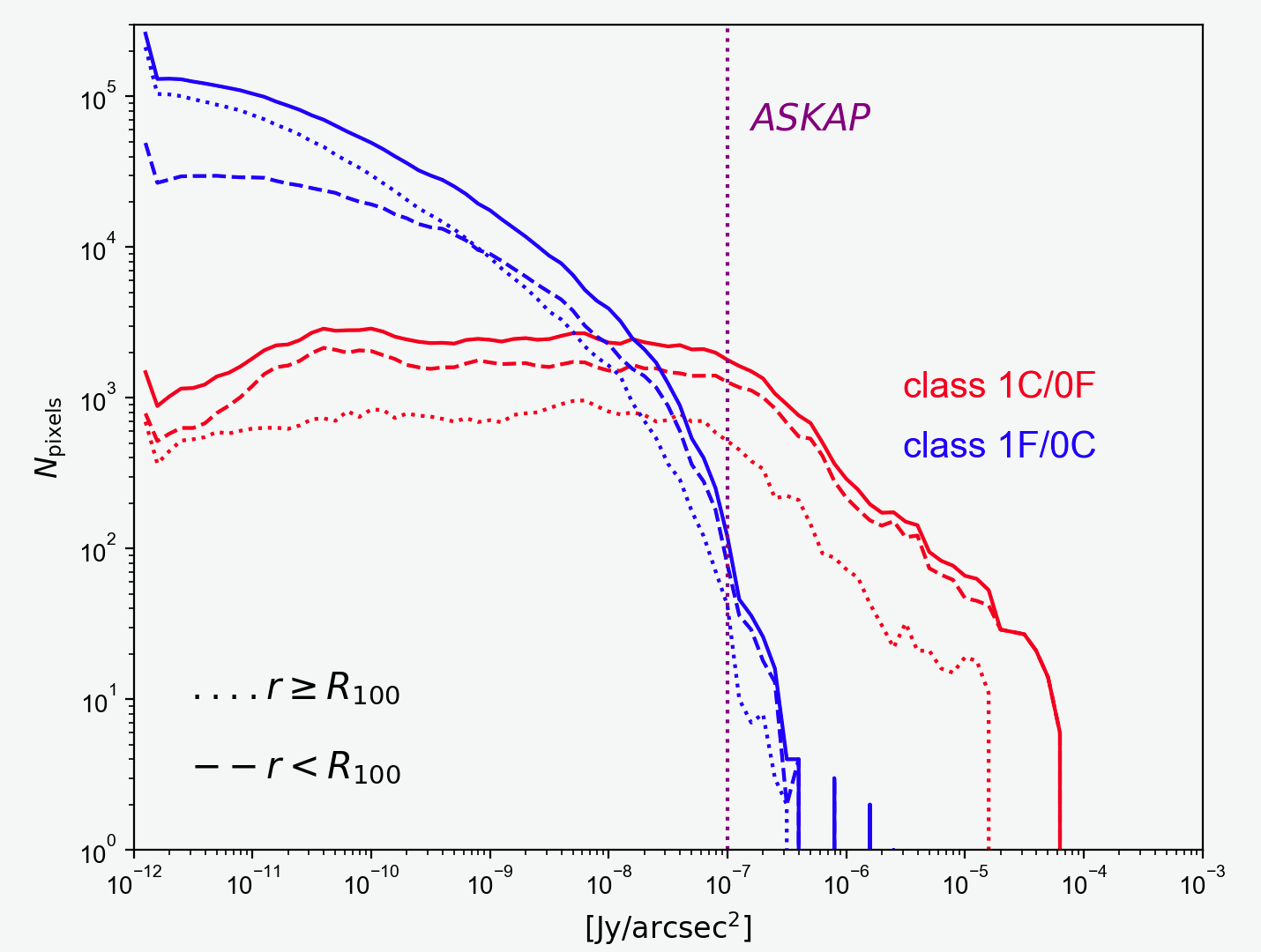}
\caption{Flux distribution in a sample 
$2000 \times 2000$ pixel image, considering a $F_{th}=0.75 \times 10^{-7} \rm Jy/arcsec^2$ threshold and $N_{pix}=9$.  The dashed lines give the distribution from pixels within the virial volume of clusters,  while the dotted lines are for pixels located outside the virial region of clusters. The solid line is the sum of the two. The dotted vertical line give the predicted rms noise of ASKAP.}
 \label{fig:pdf}
\end{center}
\end{figure}

Finally, we present in Figure \ref{fig:pdf} the statistical analysis of 
the distribution of tiles identified by {\cosmodeep}, in relation to their projected distance to galaxy clusters in the field, which we identify in a separate step with a halo finder (working in 3 dimensions).  We tentatively consider tiles falling within a $< R_{\rm 100}$ (i.e. the virial radius) from the center  of a nearby halo as ``cluster emission patches", and tiles at $\geq R_{\rm 100}$ distance as ``cosmic web emission patches".  The  tiles correctly classified by {\cosmodeep} as containing a structure have a distribution of pixel luminosities that peaks towards higher values. A very significant fraction of the structures correctly identified by {\cosmodeep} are related to shocked gas outside of the virial volume of clusters, which confirms that our technique is indeed capable of locating low  surface brightness emission regions in the peripheral regions of galaxy clusters, which trace accretion shocks and shocks around filaments. 

\section{Conclusions}
\label{sec:conclusions}

The work presented in this paper shows that a Deep Learning based methodology based on a CNN approach ({\cosmodeep}) offers an effective solution for the fully automated processing pipeline of big radio datasets, of the order of what is expected from next generations of surveys with radio telescopes (e.g. ASKAP, MEERKAT, MWA, LOFAR and the SKA). We explored the case study of extended cosmological radio sources (such as emission from shocked gas around galaxy clusters and filaments). {\cosmodeep} allows us to detect diffuse radio sources and to localize their position within large images thanks to a tiling based procedure. The overall accuracy of the method is comparable to that of more standard tools used in radio astronomy, but it delivers better performance when applied to the detection of faint objects, with emissivity below the average rms noise of radio observations.  The accuracy has been defined as the probability that a tile classified as signal contains an actual radio source. Depending on the specific set-up, accuracy is between 0.88 and 0.92, with only few tiles misclassified per 2000$\times$2000 pixel (2500 tiles) image. Such values have to be taken as a conservative lower bound, having proved that part of the observed inaccuracies are due to the tiling procedure and not to the CNN classifier itself.  

The performance of {\cosmodeep} is not only promising in terms of accuracy and ability to identify faint diffuse objects, but its computational performance is also encouraging:  2000$\times$2000 pixel images can be processed in less than $0.1$ sec on a state-of-the-art GPU, and their size does not represent an issue in terms of memory  thanks to the effective implementation provided by the TensorFlow framework. The training can be easily managed on the computing nodes used for testing, the full training on $\sim 30000-40000$ tiles requires less than one hour to be completed. Larger datasets, in terms of both image size and data volume, are potentially manageable as well, since TensorFlow supports distributed training on parallel high performance computing architectures. 

An important ``by-product'' of our methodology is a set of mock radio images generated from cosmological numerical simulations. The image set is composed of {\it Sky} images used for automatic labeling, and {\it Noise} images derived from the Sky images by adding random noise in order to create realistic radio observations. The resulting data set is unique and it can be publicly accessed at http://cosmosimfrazza.myfreesites.net/cosmodeep-training-datasets.

In summary, this  work led to the following achievements:
\begin{itemize}
\item availability of a novel methodology, based on a Deep Learning CNN approach, to detect diffuse and faint sources in radio observations, irrespectively of their specific size or shape;
\item the methodology is competitive in terms of accuracy with state-of-the-art software adopting more standard approaches;
\item the methodology is flexible and extensible to encompass a broad spectrum of applications and cases and it is scalable to increasingly bigger configurations, supporting high performance computing solutions;
\item the methodology can classify 2000$\times$2000 pixel images ``real time'' ($\sim 0.1$ sec/image)  on a state-of-the-art GPU;  
\item public availability of a datasets composed by hundreds of images generated from cosmological numerical simulations mimicking real radio observations. The datasets will be progressively extended in order to include more and more sophisticated images.
\end{itemize}

The methodology will be further developed in order to be ready for real observations, addressing in particular ASKAP data as a test-bed for the even larger challenge posed by the Square Kilometer Array. Development will progressively extend to increasingly complex and realistic images, e.g. including image processing artifacts like secondary radio lobes, remaining point-like sources and confusion noise.

\section*{Acknowledgements}
We thank Tim Dettmers for the outstanding help and support at the beginning of the work, Baerbel Koribalski for the useful discussion and suggestions and Tim Dykes for the critical reading of the paper. We acknowledge useful feedback on the manuscript by V. Luckic and M. Br\"{u}ggen.  
F.V. acknowledges financial support from the ERC  Starting Grant "MAGCOW", no. 714196.   A.B. acknowledges support from the ERC Starting Grant "DRANOEL", no. 714245. 
We acknowledge the  usage of computational resources on the Piz-Daint supercluster at CSCS-ETHZ (Lugano, Switzerland) under projects s701 and s805. 

\clearpage
\bibliographystyle{mnras}
\bibliography{sample,cosmoref}

\end{document}